\documentclass[twocolumn,showpacs,preprintnumbers,amsmath,amssymb]{revtex4}

\usepackage{graphicx}% Include figure files
\usepackage{dcolumn}% Align table columns on decimal point
\usepackage{bm}% bold math
\usepackage{graphicx}
\usepackage{epsfig}
\include{graphic}

\begin{document}
\preprint{APS/123-QED}
\title{Quantized spin excitations in a ferromagnetic microstrip from microwave photovoltage measurements}
\author{Y. S. Gui\footnote{Electronic address: ysgui@physics.umanitoba.ca}, N. Mecking and C. -M. Hu\footnote{Electronic address: hu@physics.umanitoba.ca; URL: http://www.physics.umanitoba.ca/$\sim$hu}}
\affiliation{Department of Physics and Astronomy, University of
Manitoba, Winnipeg, Canada R3T 2N2}

\date{\today}

\begin{abstract}
Quantized spin excitations in a single ferromagnetic microstrip
have been measured using the microwave photovoltage technique.
Several kinds of spin wave modes due to different contributions of
the dipole-dipole and the exchange interactions are observed.
Among them are a series of distinct dipole-exchange spin wave
modes, which allow us to determine precisely the subtle spin
boundary condition. A comprehensive picture for quantized spin
excitations in a ferromagnet with finite size is thereby
established. The dispersions of the quantized spin wave modes have
two different branches separated by the saturation magnetization.
\end{abstract}



\maketitle

Understanding quantized spin excitations in ferromagnets with
finite size is not only pivotal for exploring nanomagnetism
\cite{Tulapurkar2004}, but also essential for designing
high-density magnetic memories with fast recording speed
\cite{Prinz1998}. The most compelling topics that have recently
attracted great interest include: the interplay between
dipole-dipole and exchange interactions \cite{Jorzick1999,
Jorzick2002, Wang2002, Park2002}, the characteristics of the spin
boundary conditions \cite{Rappoport2004}, and the evolution of
spin excitations in various phases \cite{Wang2002,
Tartakovskaya2005, Wang2005}. Despite general consensus on the
theoretical explanation of the combined effects of dipole-dipole
and exchange interactions \cite{Sparks1970PRB, Kalinikos1986},
experiments found usually either magnetostatic modes (MSM)
\cite{White1956} or standing spin waves (SSW) \cite{Seavey1958},
which are determined by dipole-dipole or exchange interaction,
respectively. As a related problem, the impact of spin boundary
conditions, which has been studied over decades on thin films with
a thickness comparable to the wavelength of spin waves, remains
elusive \cite{Rado1959,Morrish2001}. The most appealing quantized
dipole-exchange spin wave (DESW) modes existing in
laterally-structured ferromagnets, which should exhibit combined
characteristics of the MSM and SSW, have only been recently
observed near the uniform ferromagnetic resonance (FMR)
\cite{Jorzick1999, Jorzick2002, Wang2002, Park2002}, and are
therefore found to be insensitive to the exchange interaction and
spin boundary conditions \cite{Sparks1970PRB}. The lack of a
comprehensive picture of spin excitations in ferromagnets with
finite size is partially due to the experimental challenge of
detecting spin waves in samples with shrinking dimensions, where
conventional techniques such as the FMR absorption and Brillouin
light scattering are approaching their sensitivity limit.

Very recently, promising new experimental techniques have been
developed for studying spin dynamics: microwave photoconductivity
\cite{Gui2005} and photovoltage techniques \cite{Gui2006}, which
allow electrical detection of spin excitations in ferromagnetic
metals. The associate high sensitivity makes it possible to
investigate the comprehensive characteristics of quantized spin
excitations.

In this letter we report investigations of quantized spin waves in
a single ferromagnetic microstrip using the microwave photovoltage
technique. Both the even and odd order SSWs are detected, and
quantized DESWs are observed near both the FMR and the SSW. Two
distinct branches of the field dispersion for the quantized spin
waves are measured. The spin boundary conditions are precisely
determined. And an empirical expression describing the dispersion
characteristics of the complete spin excitations in the entire
magnetic field range is obtained.

Our sample is a Ni$_{80}$Fe$_{20}$ (Permalloy, Py) microstrip,
with dimensions of $l$ = 2.45 mm, $w$ = 20 $\mu$m, and $d$ = 137
nm as shown in Fig. 1(a) in a $x-y-z$ coordinate system. From
anisotropic magnetoresistance measurements, we determine the
saturation magnetization $\mu_0M_0$=1.0 T. As shown in Fig. 1(b),
the Py strip is inserted in the slot of a ground-signal-ground
coplanar waveguide (CPW) made of an Au/Ag/Cr (5/550/5 nm)
multilayer. The device is deposited on a semi-insulating GaAs
substrate. By feeding the CPW with a few hundreds mW microwaves, a
d.c. voltage $V$ is measured along the $x$-axis as a function of
the magnetic field $H$ applied nearly perpendicular to the Py
strip. The photovoltage is induced by the spin rectification
effect whose characteristics are reported elsewhere
\cite{Gui2006}. The data presented here are taken by slightly
tilting the field direction away from the $z$-axis towards the
$x$-axis by a very small angle of 0.2$^{\circ}$, so that the
$x$-component of the magnetization $M_{x}$ is nonvanishing, and
the photovoltage $V\propto M_{x}$ has a power sensitivity
approaching 0.1 mV/W.

\begin{figure} [t]
\begin{center}
\epsfig{file=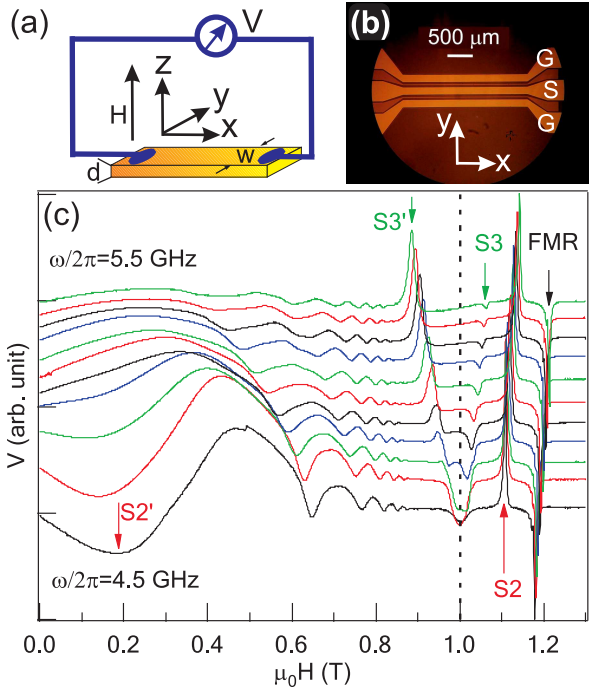,width=8cm} \caption{(color online) (a)
Schemetic drawing of the Py microstrip and the measurement
circuitry. (b) Top view micrograph of a device with Py strips
placed in slots between the ground (G) and signal (S) lines of a
coplanar waveguide. (c) Typical photovoltage spectra measured at
different microwave frequencies (from 4.5 GHz to 5.5 GHz with a
step size of 0.1 GHz). Arrows indicate FMR, SSW for the quantized
number $p$ = 2 (S2 and S2') and SSW for $p$ = 3 (S3 and S3'),
respectively. The dashed line indicates $H=M_0$. All curves are
normalized with the FMR amplitude and vertically offset for
clarity.}\label{Fig.1}
\end{center}
\end{figure}

Figure 1(c) shows the electrically-detected quantized spin
excitations in the Py microstrip. The sharp resonances at $H>M_0$
(labelled as FMR, S2 and S3) move to higher fields with increasing
microwave frequency. At high frequencies ($\omega/2\pi>8$ GHz)
another resonance (S4) is observed (not shown) at $H>M_0$. The
dispersions of these resonances follow the well-known Kittel
formula for SSWs used in textbooks \cite{Morrish2001}, given by:
$\omega=\gamma(H-M_0+2Ak_z^2/\mu_0M_0)$. The gyromagnetic ratio is
determined to be $\gamma$ = 181 $\mu_0$GHz/T. Here $k_z=(p-\Delta
p)\pi/d$ is the wave vector, and $A$ is the exchange stiffness
constant. The quantized number $p$ is the integer number of half
wavelengths along the $z$ direction. The correction factor $\Delta
p$ is bounded by $0\leq\Delta p\leq1$ and is determined by the
boundary condition \cite{Rado1959}:
\begin{equation}
2A\frac{\partial\Psi_p}{\partial z}-K_s\Psi_p=0
\end{equation}

\noindent where the eigenfunction of SSW has the form $
\Psi_p=\alpha\sin k_zz+\beta\cos k_zz$. The constants $\alpha$ and
$\beta$ are determined by both the surface anisotropy $K_s$ and
the exchange stiffness constant $A$. If
$\alpha/\beta\rightarrow\infty$, the spins at surfaces are
completely pinned and $\Delta p=0$. In the opposite case where
$\alpha/\beta\rightarrow 0$, the spins at surfaces are totally
free and $\Delta p=1$. Based on the Kittel formula, the observed
four resonances correspond to FMR ($p$ = 0) and SSWs with $p$ = 2,
3 and 4. However, the precise values of $\Delta p$, which are
dependent on $p$ in general, are difficult to deduce directly from
the resonant positions of the SSWs. This is a long standing
problem of the spin boundary condition
\cite{Rado1959,Morrish2001}, which not only sets up an obstacle
for identifying SSWs, but also causes significant diversity
\cite{Jorzick2002,Park2002,Seavey1958,Morrish2001} in determining
important spin properties such as the value of the exchange
stiffness constant $A$.

Before we proceed to determine precisely the value of $\Delta p$
by going beyond the simple Kittel picture, we briefly highlight
two interesting features observed in Fig. 1(c). One feature is
that there are two branches for each SSW modes. For example, at
$\omega/2\pi=4.5$ GHz, the $p$ = 3 SSW mode appears as a dip at $H
= M_0$. At higher frequencies, it splits into two structures: the
higher branch (dips labelled as S3) at $H > M_0$ and the lower
branch (peaks labelled as S3') at $H < M_0$. Similar effects are
observed for other SSWs (see Fig. 3 for the entire dispersions).
The higher branch is typical for the SSWs reported earlier, where
the magnetization $\mathbf{M}$ is forced to align nearly parallel
to $\mathbf{H}$, and the internal field $H_i \thickapprox H-M_0$.
The lower branch is less familiar. Here, $H_i \approx$ 0, and the
direction of $\mathbf{M}$ is tilted away from the $z$-axis towards
the $x$-axis by an angle $\varphi$ given by $cos\varphi \approx
H/M_{0}$ \cite{Gui2006}. We note that similar evolution of spin
waves observed in Ni nanowires \cite{Wang2002} and nanorings
\cite{Wang2005}, were interpreted as reorientation phase
transitions \cite{Tartakovskaya2005} and the transition from a
"twisted bamboo" state to a "bamboo" state \cite{Wang2005},
respectively.

More interestingly, Fig. 1(c) shows a series of pronounced
oscillations between S2' and S3'. The amplitude of these
oscillations decreases with increasing field strength $H$. To the
best of our knowledge, such striking oscillations, related to spin
dynamics, have never been reported before. They are observed in a
series of samples with different thickness in our experiment. As
discussed below, the oscillations originate from the lower branch
of the DESWs at $H < M_0$.

Going beyond Kittel's picture, the dispersion of DESW modes has a
form given by Kalinikos and Slavin \cite{Kalinikos1986}:
\begin{equation}
\omega^2=\gamma^2(H_i+2Ak^2/\mu_0M_0)(H_i+2Ak^2/\mu_0M_0+M_0F_p),
\end{equation}
\noindent where
\begin{eqnarray}
F_p&=&P_p+\sin^2\varphi\left(1-P_p+\frac{M_0P_p(1-P_p)}{H_i+2Ak^2/\mu_0M_0}\right),\nonumber \\
P_p&=&\frac{k_y}{2}\int_0^d\int_0^d
\Psi_p(z)\Psi_p(z')\exp(-k_y\mid z-z'\mid)dzdz'. \nonumber
\end{eqnarray}
\noindent Here $k^2=k_z^2+k_y^2$ is the wave vector, and
$k_y=n\pi/w$ is the quantized wave vector along the $y$ direction.
Neglecting the exchange effect ($A$ = 0) the DESW modes near the
FMR reduce to the magnetostatic modes with quantized number $n$.
In different measurement geometries, magnetostatic modes may
appear either as magnetostatic forward volume modes (MSFVM),
magnetostatic backward volume modes, or Damon-Eshbach modes
\cite{Spin2002}. On the other hand, neglecting the dipolar dynamic
field by assuming $P_p$ = 0, Eq. (2) reduces to the case for SSWs
with the quantized number $p - \Delta p$.

\begin{figure} [t]
\begin{center}
\epsfig{file=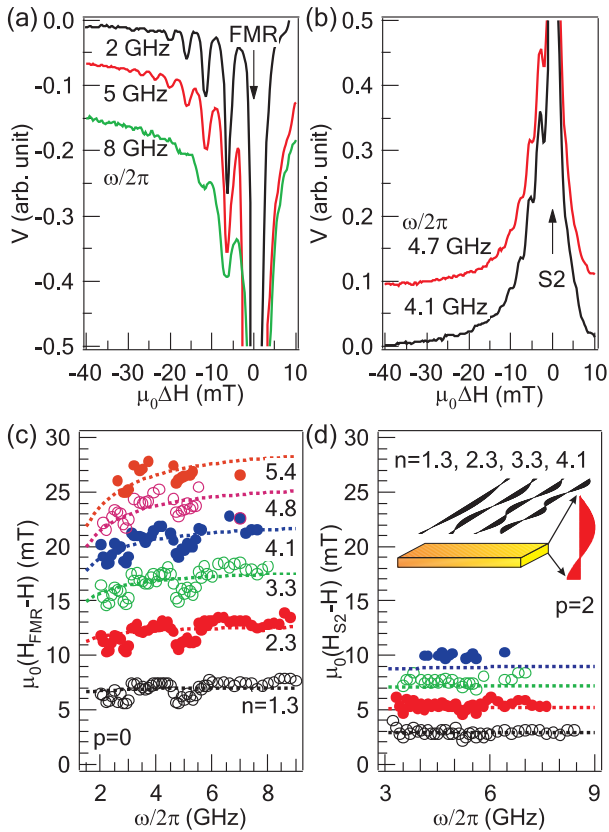,width=8 cm} \caption{(color online) (a)
Quantized MSFVM and (b) DESW modes found near FMR ($p$=0) and SSW
($p$=2), respectively. The spectra are measured at different
microwave frequencies and are vertically offset. They are
normalized either to FMR or SSW ($p$=2). (c) The measured
dispersions (symbols) of the quantized MSFVM and (d) DESW modes
are compared with the calculated results (dotted lines). The inset
in (d) illustrates the spin waves in the microstrip quantized
along both $y$ and $z$ directions. }\label{Fig.2}
\end{center}
\end{figure}

Both MSFVMs and SSWs are detected by the photovoltage technique in
our experiment. If we focus on the low-field range of the FMR, a
series of fine structures are well resolved as shown in Fig. 2(a).
These are the quantized MSFVMs. The first MSFVM has an intensity
of about 25\% of that of the FMR and its width is narrower but
comparable to that of the FMR (a few mT). The intensity of the
MSFVM dramatically decreases with increasing $n$, while its width
is not sensitive to $n$. The widths of both the FMR and the MSFVMs
increase with microwave frequency roughly following a linear
relation due to Gilbert damping \cite{Heinrich2004}. Using
$\Psi_0=1/\sqrt{d}$ one obtains $P_0 \sim k_yd/2 \sim$ 10$^{-2}n$
in the long-wavelength limit ($k_yd\ll1$) for MSFVMs.
Consequently, the dispersions of the quantized MSFVMs are
essentially independent of both the boundary conditions and the
exchange interaction, as also pointed out by Sparks
\cite{Sparks1970PRB}. Fig. 2(c) shows the resonance positions of
the MSFVMs (symbols) as a function of the microwave frequency. The
dotted lines are calculated according to Eq. (2) by adjusting the
quantized number $n$. The resulting values of $n$ are 1.3, 2.3,
3.3, 4.1, 4.8, and 5.4. The spacing between the MSFVM and the FMR
saturates at a value of $P_0M_0/2$ at high frequencies when
$\omega\gg \gamma P_0M_0$.

The significance of this work is observing not only both the
quantized MSFVMs and SSW modes, but also a distinct type of
quantized DESW mode determined by both the quantized numbers $n$
and $p-\Delta p$; here the interplay between the exchange and
dipole-dipole interactions is significant, and the surface spin
pinning must be taken into account. For $H > M_{0}$, as shown in
Fig. 2(b), the quantized DESW modes ($p$ = 2, $n \neq$ 0) appear
as a series of discrete resonances on the lower field side of the
SSW with $p = 2$. The spacing between these modes is of the same
order of magnitude as the MSFVMs. This implies that the expression
for $P_{2}$ is similar to $P_{0}$ and may be of the form $k_yd$.
Indeed, Fig. 2(d) shows good agreement between the measured
dispersions and the calculated results using Eq. (2) with $P_2 =
2k_yd/\pi^2 \sim 4 \times 10^{-3}n$. The quantized numbers $n$ are
the same values as those for the quantized MSFVMs. It should be
emphasized that the theoretical expression for $P_{2}$ depends
strongly on the spin boundary conditions. For totally unpinned
surface spins, one obtains $P_p=(k_yd/p\pi)^2$. For totally pinned
surface spins, $P_p=3(k_yd/p\pi)^2$ for even $p$, and
$P_p=4k_yd/p^2\pi^2+3(k_yd/p\pi)^2$ for odd $p$. In order to
explain the observed DESW modes near the SSW with $p = 2$, unequal
spin pinning at two surfaces of the Py microstrip must be taken
into account. Here we assume that the spins are fully pinned only
at the top surface by a thin antiferromagnetic oxide layer there,
while the spins are partially pinned (described by $\Delta p$) at
the bottom surface adjacent to the GaAs substrate. Using the
experimental value of $P_2=2k_yd/\pi^2$, we deduce $\Delta p$ ($p$
= 2) to be 0.75. By using both Eqs. (1) and (2), we further
determine $K_s\sim8\times10^{-4}$ N/m and $A=1.4\times10^{-11}$ N
from the measured dispersion for the SSW with $p = 2$. Then, the
values of $p-\Delta p$ for other SSWs are deduced from Eq. (1) to
be 0, 1.25, 2.35, and 3.4 for the SSWs with $p = $1, 2, 3 and 4.
Note that the SSW for $p=1$ determined under such a spin boundary
condition coincides with the FMR as found in the experiment, and
the observed four resonances at $H>M_{0}$ are identified as FMR
($p$ = 0) and SSWs with $p$ = 2, 3, and 4. The calculated
intensities of SSWs based on such a spin boundary condition are in
good agreement with the experimental results: The intensities of
the FMR and the SSW ($p = 2$) are comparable and are both much
stronger than the intensities of higher order SSWs, while the
intensity of the SSW ($p$ = 4) is always stronger than that of the
SSW ($p$ = 3). Additionally, $P_3$ is calculated to be about
0.05$k_yd/\pi^2 \sim 10^{-4}n$, much smaller than $P_2$. This
explains the result that DESW modes have been observed near
neither branch of SSW with $p = 3$.

\begin{figure} [t]
\begin{center}
\epsfig{file=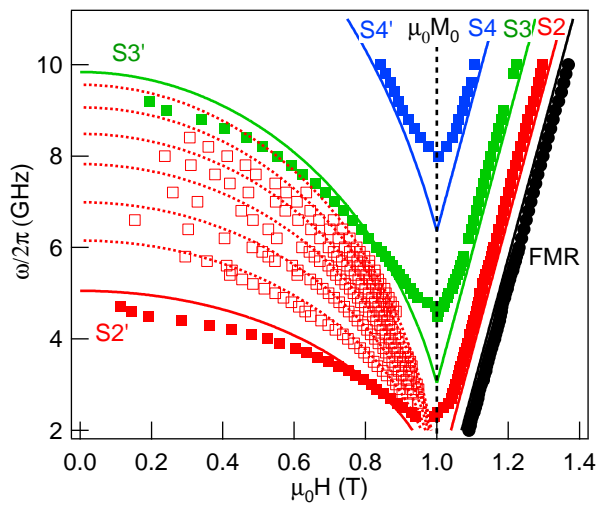,width=8 cm} \caption{(color online)
Dispersions of FMR ($p$ = 0) and SSWs ($p$ = 2 for S2 and S2', $p$
= 3 for S3 and S3', $p$ = 4 for S4 and S4') measured (solid marks)
in the entire magnetic field range. Open symbols show the measured
lower branches of the DESWs at $H < M_0$. Upper branches at $H >
M_0$, which are too close to the FMR and SSW, are not shown here
for brevity but are plotted in Fig. 2 instead. Solid lines are
calculated for the SSWs with $p$ = 0, 2, 3 and 4. Dotted lines are
calculated for the DESWs with $p$ = 2. The dashed line indicates
$H = M_0$.}\label{Fig.3}
\end{center}
\end{figure}

The precisely determined spin boundary conditions allow us to
establish a complete picture for the quantized spin excitations.
Figure 3 shows the dispersions of the quantized spin waves in the
entire field range. At $H>M_{0}$, the solid symbols labelled as
FMR, S2, S3 and S4 are FMR ($p$ = 0) and SSWs with $p$ = 2, 3 and
4. Note that the SSWs evolve into the S2', S3' and S4' at
$H<M_{0}$. At $H>M_{0}$, the fine structures of the quantized
MSFVMs ($p=0$) and DESW ($p=2$), which appear at the lower field
side of the FMR and S2, respectively, are not shown in Fig. 3 for
clarity. Their dispersions are displayed in Fig. 2 instead.
Oscillations between S2' and S3' found in Fig. 1(c) at $H < M_{0}$
can now be understood as modes that evolved from the quantized
DESW ($p=2$) near S2 at $H>M_{0}$. Resonance positions at the
minima of these oscillations are displayed by the open symbols in
Fig. 3. We obtain an empirical expression describing the complete
spin wave modes with the quantized numbers ($p-\Delta p$, $n$) in
the entire field range:
\begin{eqnarray}
\omega^2&=&\gamma^2(H_i+2Ak^2/\mu_0M_0+P_pM_0)(H_i+2Ak^2/\mu_0M_0\nonumber \\
&&+M_0(1+2n/\pi^2)(1-P_p)\sin^2\varphi)
\end{eqnarray}

\noindent where $P_0=k_yd/2$ and $P_2=2k_yd/\pi^2$. Results
calculated (curves) using Eq. (3) agree well with experimental
data \cite{explaination1}. Note that by assuming $n=0$, Eq. (3)
reduces to Eq. (2) describing SSWs in the entire $H$ range, and by
assuming $\varphi = 0$, it agrees with Eq. (2) for the DESW modes
at $H>M_0$.

In summary, using a highly sensitive photovoltage technique, a
comprehensive picture of quantized spin excitations in a single Py
microstrip is established. The characteristics of a distinct
series of DESW modes allow us to determine precisely the spin
boundary condition. The results pave a new way for studying spin
dynamics in ferromagnets with finite size, where both the
geometrical effect and spin boundary conditions play important
roles.

We thank B. Heinrich, G. Williams, H. Kunkel and X. Z. Zhou for
discussions, G. Roy for technical support, D. Heitmann, U. Merkt
and DFG for the loan of an equipment. N. M. is supported by a DAAD
scholarship. This work has been funded by NSERC and URGP.


\begin{thebibliography}{100}
\bibitem {explaination1} The discrepancy around $H = M_0$ is
caused by the simplified calculation of $H_i$, where we assume a
perpendicular field direction by neglecting the small tilting
angle of 0.2$^{\circ}$. At $H = M_0$, this induces an deviation of
36.5 mT for $H_i$, which causes a resonance frequency discrepancy
of about 1.05 GHz.
\end{thebibliography}
\end{document}